\documentclass[conference]{IEEEtran}
\IEEEoverridecommandlockouts
\usepackage{array, graphicx, float}
\usepackage{stix}
\usepackage{amsmath}
\usepackage{ragged2e}
\usepackage{titlesec}
\usepackage{booktabs}
\usepackage{fancyhdr}
\usepackage{soul}
\usepackage[english]{babel}
\usepackage[style=ieee, backend=biber, minbibnames=2, maxbibnames=2]{biblatex}

\usepackage{csquotes}
\usepackage{rotating}
\usepackage[normalem]{ulem}
\usepackage{enumitem}
\usepackage{textgreek}
\usepackage{listings}
\usepackage{lscape}
\usepackage{caption}
\DeclareMathOperator{\Tr}{Tr}

\usepackage{tabularx}
\newcolumntype{h}{>{\hsize=2.0\hsize}X}
\newcolumntype{e}{>{\hsize=1.75\hsize}X}
\newcolumntype{g}{>{\hsize=1.5\hsize}X}
\newcolumntype{l}{>{\hsize=1.25\hsize}X}
\newcolumntype{b}{>{\hsize=1.15\hsize}X}
\newcolumntype{x}{X}
\newcolumntype{n}{>{\hsize=.85\hsize}X}
\newcolumntype{m}{>{\hsize=.75\hsize}X}
\newcolumntype{s}{>{\hsize=.5\hsize}X}
\newcolumntype{t}{>{\hsize=.25\hsize}X}

\usepackage[colorinlistoftodos]{todonotes}

\begin{document}
\captionsetup[figure]{name={Fig.}}
\setlength{\abovedisplayskip}{6pt}
\setlength{\belowdisplayskip}{6pt}

\onecolumn
\null
\vfill
\begin{center}
    \huge{This article has been accepted for publication in the IEEE International Conference on Quantum Communications, Networking, and Computing 2025. This is the accepted manuscript made available via arXiv. }
\end{center}
\vfill
\normalsize{© 2025 IEEE. Personal use of this material is permitted. Permission from IEEE must be obtained for all other uses, in any current or future media, including reprinting/republishing this material for advertising or promotional purposes, creating new collective works, for resale or redistribution to servers or lists, or reuse of any copyrighted component of this work in other works.}

\twocolumn
\clearpage

\title{Machine Learning for Phase Estimation in Satellite-to-Earth Quantum Communication}

\author{Nathan~K.~Long$^{*\dagger}$, 
        ~Robert~Malaney$^*$, 
        ~Kenneth~J.~Grant$^\dagger$
\thanks{$^*$Nathan K. Long and Robert Malaney are with the School of Electrical Engineering and Telecommunications, University of New South Wales, Kensington, NSW, Australia. $^\dagger$Kenneth J. Grant was with the Sensors and Effectors Division of Defence Science and Technology Group, Edinburgh, SA, Australia ($^\dagger$NKL is currently on secondment from this Division).}
}

\maketitle

\begin{abstract}
    A global continuous-variable quantum key distribution (CV-QKD) network can be established using a series of satellite-to-Earth channels. Increased performance in such a network is provided by performing coherent measurement of the optical quantum signals using a real local oscillator, calibrated locally by encoding known information on transmitted reference pulses and using signal phase error estimation algorithms. The speed and accuracy of the signal phase error estimation algorithm are vital to practical CV-QKD implementation. Our work provides a framework to analyze long short-term memory neural network (NN) architecture parameterization, with respect to the quantum Cram{\'e}r-Rao uncertainty bound of the signal phase error estimation, with a focus on reducing the model complexity. More specifically, we demonstrate that signal phase error estimation can be achieved using a low-complexity NN architecture, without significantly sacrificing accuracy. Our results significantly improve the real-time performance of practical CV-QKD systems deployed over satellite-to-Earth channels, thereby contributing to the ongoing development of the Quantum Internet.
\end{abstract}

\section{Introduction}

Quantum key distribution (QKD) utilizes the principles of quantum mechanics to share a set of keys between two parties for information-theoretic unconditional security. The use of a low Earth orbit satellite-based network to distribute keys between parties has the potential to enable global security using QKD. Continuous-variable QKD (CV-QKD), using Gaussian-modulated coherent states, involves encoding key information on the quadratures of an electric field, with the advantage of being operational with classical optical hardware and digital signal processing, when compared to discrete-variable QKD~\cite{Chen2023}. Both homodyne and heterodyne detectors can be used for coherent measurement of CV quantum signals (hereafter, referred to simply as ``signals'')~\cite{Grosshans2002}.

In the context of free-space optical CV-QKD, transmitted local oscillators (TLOs) were traditionally considered for coherent measurement of the signals. Current state-of-the-art CV-QKD protocols, however, consider using a real local oscillator (RLO) created locally at the receiver, mitigating security loopholes associated with TLOs~\cite{Ma2013}. At the receiver, the RLO is manipulated so as to correct for the phase errors accumulated across the channel, such errors being measured from reference pulses that are polarization multiplexed with the signals. However, there remains a difference in the phase error measured using the reference pulses and the phase error of the signals, the reduction of which using signal phase error estimation algorithms is an ongoing area of research~(e.g.~\cite{Soh2015, Chin2021, Zhang2023}). Given that operational speed remains a primary factor in designing CV-QKD systems to function in real-time, signal phase error estimation algorithms should be as low-complexity as possible (whilst achieving the required accuracy).

Machine learning (ML) is able to adapt to noisy data by learning patterns between input and output data, where no assumptions-based physical models are required. Previous works have used a variety of ML algorithms for signal phase error estimation~\cite{Long2023_survey}, including convolutional neural networks (NNs)~\cite{Xing2022}, long short-term memory (LSTM) NNs~\cite{Zhang2023}, and Bayesian inference with Kalman filters~\cite{Chin2021}. However, a question that remains unexplored within the context of ML applied to signal phase error estimation is: what is the lowest complexity (highest operational speed) NN that can achieve a required accuracy? This is the main question we answer here.

Our contributions can be summarized thus:
(i)~We analyze the relationship between LSTM NN model architecture and signal phase error estimation accuracy for satellite-to-Earth channels, with a focus on extracting an efficient model for the required phase error estimation, while limiting the number of free parameters within the NN (thereby offering lower complexity). (ii)~We compare several architectures, their estimation performance, and how their estimations relate to formal uncertainty bounds, calculated using Fisher information. 
(iii)~We show how our results lead to a feasible ML-based CV-QKD system for satellite-to-Earth channels that have improved outcomes relative to standard, non ML-based systems. 

The remainder of this paper is as follows:
The system model is presented in Section~\ref{sec:system}, the NN architectures are outlined in Section~\ref{sec:nn}, the NNs are analyzed in Section~\ref{sec:nn_anal}, and concluding remarks are made in Section~\ref{sec:conc}.

\section{System Model} \label{sec:system}

In our RLO-based CV-QKD protocol, Alice (subscript $A$) at the satellite encodes the reference pulses (subscript $R$) with quadrature values $\{X_{A_R}, P_{A_R}\}$, and signals (subscript $S$) with quadrature values $\{X_{A_S}, P_{A_S}\}$. The reference pulses and signals are then polarization multiplexed and transmitted across an atmospheric channel to Bob (subscript $B$) at the ground station. The encoded reference pulse information is shared with Bob beforehand, while the signal information is unknown to Bob and contains the secure key. At the receiver, Bob demultiplexes the reference pulses and signals, measuring their quadrature values $\{X_{B_R}, P_{B_R}\}$ and $\{X_{B_S}, P_{B_S}\}$, respectively. 

The encoded quadratures are distorted by the channel, resulting in a \textit{reference pulse phase error} $\Delta\phi_R$, calculated as,
\begin{equation}\label{eq:dphi_r}
    \Delta\phi_R = \tan^{-1} \left( \frac{P_{B_R} X_{A_R} - X_{B_R} P_{A_R}}{X_{B_R} X_{A_R} + P_{B_R} P_{A_R}} \right).
\end{equation}

\noindent Likewise, the signal is also distorted by the channel and detector, accumulating a \textit{signal phase error} $\Delta\phi_S$ (calculated using the same method as Eq.~\ref{eq:dphi_r}).

In our scheme, an RLO at the receiver is split into the RLO$_1$ and RLO$_2$, where a heterodyne detector uses the RLO$_1$ to measure the reference pulse quadratures, such that $\Delta\phi_R$ can be derived. Phase shifting the RLO$_2$ by $\Delta\phi_R$ for each pulse provides a first approximation of the true signal phase error $\Delta\phi_S$, $\Delta\phi_R \approx \Delta\phi_S$, increasing the coherence between the signal and RLO$_2$. However, we assume that $\Delta\phi_R$ measured by mixing the reference pulses and RLO$_1$ does not correspond exactly to $\Delta\phi_S$ measured by mixing the signals and RLO$_2$, such that ${\Delta\phi_R \neq \Delta\phi_S}$ (shown experimentally in~\cite{Zhang2023, Chin2021}).

\begin{figure}
    \begin{centering}
    \includegraphics[scale=0.75]{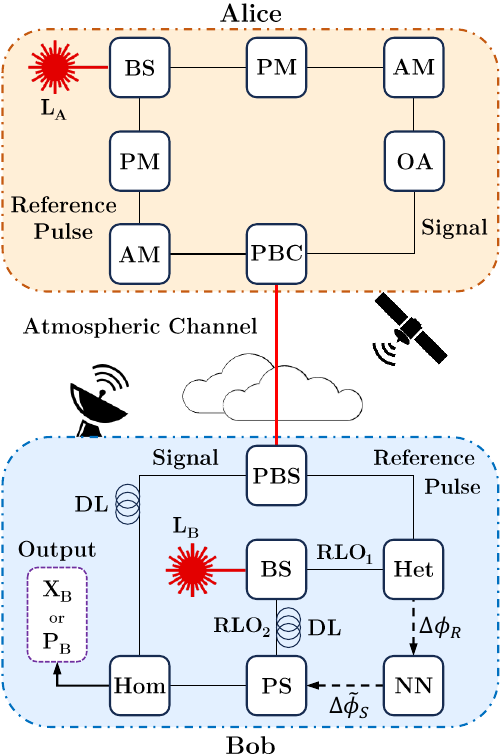} 
    \caption{Schematic outlining the preparation and measurement of coherent states with $\Delta\phi_S$ estimation. L is a laser source, BS is a beam splitter, AM is an amplitude modulator, PM is a phase modulator, OA is an optical attenuator, PBC is a polarized beam combiner, PBS is a polarized beam splitter, Het is a heterodyne detector, Hom is a homodyne detector, PS is a phase shifter, DL is a delay line, and~NN is neural network.}\label{fig:circ}
    \end{centering}
\end{figure}

Estimation of the true values of $\Delta\phi_S$ is, thus, an important aspect of RLO-based CV-QKD, where a perfect estimation of $\Delta\phi_S$, used to phase shift the RLO$_2$, would result in ideal recovery of the encoded quadrature information, ${\{X_{B_S}, P_{B_S}\} \approx \{X_{A_S}, P_{A_S}\}}$. Estimation of $\Delta\phi_S$ is undertaken using a \textit{$\Delta\phi_S$~estimation algorithm}. In practice, a $\Delta\phi_S$~estimation algorithm may be trained using known $\Delta\phi_S$ values; however, when being implemented for real-time CV-QKD, the $\Delta\phi_S$ is inherently unknown, thus only $\Delta\phi_R$ can be used to output an estimate $\Delta\Tilde{\phi}_S$ of $\Delta\phi_S$. Fig.~\ref{fig:circ} illustrates our RLO-based CV-QKD scheme, where $\Delta\phi_R$ measurements are input into a NN-based $\Delta\phi_S$~estimation algorithm (described in Section~\ref{sec:nn}), which outputs an estimate $\Delta\Tilde{\phi}_S$, used to phase shift the RLO$_2$.

Previous works~\cite{Marie2017, Shao2021, Soh2015, Kish2021} have described the relationship between $\Delta\phi_R$ and $\Delta\phi_S$ as being represented by a Gaussian function, which applies to every measurement of the reference pulses and signals. We term this the \textit{Gaussian phase noise}, which is related to the reference pulse and signal measurements being shot-noise limited (representing the standard quantum limit), as well as experiencing different interactions with the atmospheric channel and receiver hardware.

Considering the Gaussian phase noise, simply taking the expectation of the previous $N$ measurements of $\Delta\phi_R$ can result in a more precise estimate of the current $\Delta\phi_S$ value, which we define as the \textit{expectation estimator} $\langle \Delta\phi_R \rangle_N$. However, the relationship between $\Delta\phi_R$ and $\Delta\phi_S$ is complex, involving many environmental and system-based factors. Therefore, a $\Delta\phi_S$~estimation algorithm should be able to adapt to \textit{any} relationship between $\Delta\phi_R$ and $\Delta\phi_S$, such that $\Delta\phi_S$ can be estimated using the information from the $\Delta\phi_R$ measurements. Therefore, we propose a NN-based $\Delta\phi_S$~estimation algorithm capable of adapting to varying relationships between $\Delta\phi_R$ and $\Delta\phi_S$, including for the Gaussian phase noise. Note that any change in channel conditions will result in initial poor $\Delta\phi_S$ estimation performance while the $\Delta\phi_S$~estimation algorithm adapts to the new conditions.

To demonstrate the robustness of our $\Delta\phi_S$ estimation approach, we present two different relationships between $\Delta\phi_R$ and $\Delta\phi_S$. Case~1 relates to the Gaussian phase noise (Section~\ref{sec:case_1}), and Case~2 includes both the Gaussian phase noise of Case~1 \textit{and} an additional source of error between $\Delta\phi_R$ and $\Delta\phi_S$, which remains constant for each coherence time (Section~\ref{sec:case_2}). To demonstrate the advantage of our NN-based approach, we compare the $\Delta\phi_S$ estimation performance of the NNs to the approximation $\Delta\phi_R \approx \Delta\phi_S$ and the expectation estimator $\langle \Delta\phi_R \rangle_N$.

\subsection{Phase Error Simulation} \label{sec:rpe}

Previous works have described several potential sources contributing to the Gaussian phase noise~\cite{Marie2017, Shao2021, Soh2015, Kish2021}, which we adopt in our work. Specifically, we assume that the total Gaussian phase error variance $\sigma^2_{error}$ is calculated as, 
\begin{equation}\label{eq:v_est}
    \sigma^2_{error} = \sigma^2_{det} + \sigma^2_{drift}.
\end{equation}

\noindent Here, the detector phase error variance $\sigma^2_{det}$ is introduced by the shot-noise limited and experimental noise of the reference pulse heterodyne detection~\cite{Soh2015, Kish2021}, which is calculated as~\cite{Kish2021},
\begin{equation}\label{eq:v_error}
    \sigma^2_{det} = \frac{\xi_{ch} + 2\frac{1 + \xi_{det}}{\eta_{det} T}}{|\alpha_R|^2},
\end{equation}

\noindent where it is assumed that $|\alpha_R|^2$, $\xi_{ch}$, and $\eta_{det}$ are constant (defined in Table~\ref{tab:noise_params}). The transmissivity $T$ is calculated using a satellite-to-Earth phase screen model (outlined in Section~\ref{sec:channel}) for the channel conditions defined in Table~\ref{tab:sim_params}. When considering the detector noise term $\xi_{det}$ in Eq.~\ref{eq:v_error}, an additional error arises from the distortion of the reference pulse 2-dimensional phase wavefront across the channel. The coherent efficiency $\gamma$ quantifies the coherence between the reference pulse and RLO$_1$ wavefronts, where 0 represents no coherence and 1 represents full coherence. Detector noise $\xi_{det}$ is calculated using the coherent efficiency as,
\begin{equation}\label{eq:xi_det}
    \xi_{det} = \frac{((1-\gamma) + \xi_{el}) \eta_{det}}{\gamma},
\end{equation}

\noindent where the coherent efficiency is sampled from the PDF $\mathcal{N}(0.8430, 0.0025)$. See~\cite{Wang2019_coherence} for a more in-depth explanation of the coherent efficiency.

\begin{table}[htbp]
    \begin{center}
        \caption{Noise parameters.}
        \label{tab:noise_params}
        \begin{tabularx}{\linewidth}{ 
            >{\raggedright\arraybackslash}l
            | >{\centering\arraybackslash}m}
            \hline
            \textbf{Parameter} & \textbf{Value} \\ \hline \hline
            Reference pulse intensity ($|\alpha_R|^2$) & 20$V_{mod}$~\cite{Marie2017} \\ \hline
            Detector efficiency ($\eta_{det}$) & 95$\%$~\cite{Kish2021}  \\ \hline 
            Detector electronic noise ($\xi_{el}$) & 0.0100 SNU~\cite{Wang2018} \\ \hline
            Channel noise ($\xi_{ch}$) & 0.0172 SNU~\cite{Kish2021} \\ \hline
            Phase drift error variance ($\sigma^2_{drift}$) & 0.1012 SNU~\cite{Kish2021, Marie2017} \\ 
        \end{tabularx}
    \end{center}
\end{table}

Phase drift error variance $\sigma^2_{drift}$ is associated with both time-of-arrival fluctuations across the channel and imperfections in delay lines at the receiver~\cite{Marie2017, Kish2021}, defined in Table~\ref{tab:noise_params}. Note that the choice of noise values in Table~\ref{tab:noise_params} are selected from previous works~\cite{Marie2017, Kish2021, Wang2018}. While it is assumed that these values would fluctuate between CV-QKD setups, they provide a reasonable representation for the purpose of our work.

\subsection{The Satellite-to-Earth Channel} \label{sec:channel}

While previous works have investigated $\Delta\phi_S$ estimation for optical fiber channels~\cite{Long2023_survey}, free-space optical satellite-to-Earth connections remain vital to the establishment of a global QKD network.

A phase screen model based on~\cite{Beck2016} simulates the effects of the atmospheric channel on the reference pulses and signals, modified to account for the variation of atmospheric conditions as a function of altitude for the satellite-to-Earth channel. The atmosphere is conceptually separated into layers (as described in~\cite{Andrews2005}), where the width of each layer is varied as a function of altitude. The phase screens are placed at the center of each layer to simulate the distortion of the intensity and phase of the beam wavefront as if it had passed through an equivalent atmospheric volume.

To model turbulence, the refractive index structure parameter of turbulence $C^2_n$ is calculated as a function of altitude $h$ in m, root mean square wind speed $v_{rms}$ in m/s, and ground-level $C^2_n$ in m$^{-2/3}$, $C^2_n(0)$ as~\cite{Andrews2005},
\begin{equation}\label{eq:cn2}
    \begin{split}
        C^2_n(h) &= (0.00594 (v_{rms}/27)^2 \ (h \times 10^{-5})^{10} \ \exp(-h/1000) \\ 
        & + 2.7 \times 10^{-16} \ \exp(-h/1500) + C^2_n(0) \exp(-h/100)),
    \end{split}
\end{equation}

\noindent where the root mean square wind speed is given by,
\begin{equation}\label{eq:wind_rms}
    v_{rms} = \left[ \frac{1}{15 \times 10^3} \int_{5 \times 10^3}^{20 \times 10^3} v^2(h) \ dh \right]^{1/2},
\end{equation}

\noindent with $v(h)$ representing altitude-dependent wind speed in m/s. The Bufton wind profile~\cite{Andrews2005} is adopted to simulate $v(h)$, derived as a function of altitude as,
\begin{equation}\label{eq:wind}
    v(h) = v_g + 30 \exp \left[ -\left( \frac{h - 9400}{4800} \right)^2 \right],
\end{equation}

\noindent where $v_g$ is the ground-level wind speed in m/s. Note that slew rate is not considered. The system parameters for our simulations are defined in Table~\ref{tab:sim_params}.

\begin{table}[htbp]
    \begin{center}
        \caption{Simulation parameters.}
        \label{tab:sim_params}
        \begin{tabularx}{\linewidth}{ 
            >{\raggedright\arraybackslash}b
            | >{\centering\arraybackslash}n}
            \hline
            \textbf{Parameter} & \textbf{Value} \\ \hline \hline
            Satellite altitude ($H$) & 500 km \\ \hline
            Receiver altitude ($h_0$) & 2.00 km \\ \hline
            Outer scale of turbulence ($L_0$) & 5.00 m \\ \hline
            Inner scale of turbulence ($l_0$) & 0.025 m \\ \hline
            Satellite zenith angle ($\theta_z$) & $0^\circ$ \\ \hline
            Laser wavelength ($\lambda$) & 1550 nm \\ \hline
            Beam waist radius ($w_0$) & 0.150 m  \\ \hline
            Receiver radius ($R_r$) & 1.25 m \\ \hline
            Pulse transmission rate ($\mathcal{T}_R$) & 1 MHz \\ \hline
            Modulation variance ($V_{mod}$) & 10 SNU \\ \hline
            Ground-level wind speed ($v_g$) & 2.3 - 5.0 m/s \\ \hline 
            Ground-level $C^2_n$ $\left(C^2_n(0)\right)$ & {$1.70 \times 10^{-15}$} - $10^{-14}$ m$^{-2/3}$  \\
        \end{tabularx}
    \end{center}
\end{table}

Coherence time $\tau$ represents the duration of time for which channel conditions can be considered as constant. For satellite-to-Earth atmospheric channels, the coherence time is modeled as a function of $C^2_n(h)$ (Eq.~\ref{eq:cn2}), the transverse wind profile $v(h)$ (Eq.~\ref{eq:wind}), and optical wavelength $\lambda$ in m, and is calculated as~\cite{Tokovinin2008},
\begin{equation}\label{eq:coh_time}
    \tau = \left( 118\lambda^{-2} \int_{0}^{+\infty} C^2_n(h) \ v^{5/3}(h) \ dh \right)^{-3/5},
\end{equation}
\noindent 
where $v_g$ and $C^2_n(0)$ are randomly sampled from uniform distributions (defined in Table~\ref{tab:sim_params}) to generate a phase screen model for each coherence time. Transmissivity is calculated from the results as $T =(P_R/P_T)$, where $P_T$ and $P_R$ represent the power at the transmitter and receiver, respectively.

\section{Neural Network Architecture} \label{sec:nn}

\subsection{Long Short-Term Memory Model}

The $\Delta\phi_S$~estimation algorithm needs to map the time-varying relationship between $\Delta\phi_R$ and $\Delta\phi_S$, estimating $\Delta\phi_S$ for each pulse by taking $N$ $\Delta\phi_R$ measurements as input. An LSTM NN is a type of recurrent NN, which incorporates both long and short-term memory components. We provide only a very brief overview here of LSTM NNs, for a detailed description see~\cite{Yu2019}. The $N$ previous $\Delta\phi_R$ measurements can be used to output an estimate, $\Delta\Tilde{\phi}_S$, of $\Delta\phi_S$ for the current pulse, where $N$ also defines the number of LSTM units within the NN. An LSTM unit is a structure that contains three different gate types - the combination of which control the information stored in the long-term and short-term  memories. Normally, the long-term and short-term memories are abstracted as the vectors $c_t$ and $h_t$, respectively, both vectors of length $z_{dim}$~\cite{Yu2019}.

\begin{figure}
    \begin{centering}
    \includegraphics[scale=1.27]{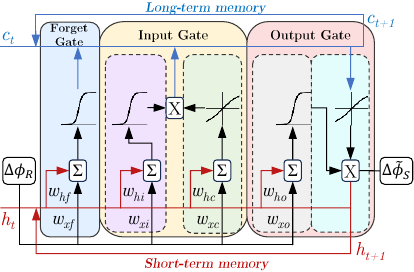} 
    \caption{LSTM unit architecture, showing the forget, input and output gates. Summations are represented by $\mathrm{\Sigma}$, dot-products by $\mathrm{X}$, and the time steps are defined by subscript $t$. Sigmoid and tanh activation functions are shown within the gates.}\label{fig:lstm}
    \end{centering}
\end{figure}

A schematic of the LSTM unit designed for this work is given in Fig.~\ref{fig:lstm}. In Fig.~\ref{fig:lstm}, $w_{xf}$, $w_{xi}$, $w_{xc}$, and $w_{xo}$ are weight vectors of length $z_{dim}$, and $w_{hf}$, $w_{hi}$, $w_{hc}$, and $w_{ho}$ are weight matrices of dimension ${z_{dim} \times z_{dim}}$. During run-time, each element of the weight vectors is multiplied by $\Delta\phi_R$, and each element of the weight matrices is multiplied by $h_t$. For each pulse, the measurement $\Delta\phi_R$ is input into the unit. The Forget Gate multiplies each value in $c_t$ by a number between 0 and 1, where the number is determined via a sigmoid activation function that takes as input $w_{xf}$ and $w_{hf}$. The Input Gate updates $c_t$ in a two-step process involving both a sigmoid and the tanh function, as illustrated. The Output Gate updates $h_t$, also through a two-step process (as illustrated), except with the incorporation of information from $c_t$. The Output Gate also passes $c_t$ and $h_t$ to the next time step, and outputs an estimate $\Delta\Tilde{\phi}_S$ via a fully-connected layer. Essentially, the NN updates itself in real-time to the previous $N$ measurements of $\Delta\phi_R$, learning long-term and short-term dependencies of the fluctuating relationship with $\Delta\phi_S$. The output estimate $\Delta\Tilde{\phi}_S$ for each pulse is then used to phase shift the RLO$_2$ for coherent measurement of the signal.

\subsection{Scale and Training}

Given that only $z_{dim}$ and $N$ are varied between NN models, we compare their complexity in terms of the number of free parameters in the model (e.g. such as in~\cite{Tan2019}), where we define the NN architecture scale $S$ as ${S = 4N\left[(z_{dim} + 1)z_{dim} + z_{dim}\right]}$~\cite{Yu2019}.

All NNs are trained using approximately a million measurements of input $\Delta\phi_R$ and output $\Delta\phi_S$, representing approximately a hundred coherence times, where $\Delta\phi_S$ is sampled from the PDF $\mathcal{N}(0.2618$, $0.0524$) for each coherence time, and each measurement of $\Delta\phi_R$ is calculated using the phase errors defined in Section~\ref{sec:rpe}. The trained NNs are then tested on approximately a million input $\Delta\phi_R$ measurements, outputting an estimate $\Delta\Tilde{\phi}_S$ for each pulse.

\subsection{Quantum Cram{\'e}r-Rao Bound} \label{sec:qcrb}
In testing the performance of our NNs it will be useful to consider the optimal performance, which is attained via the quantum Cram{\'e}r-Rao bound (QCRB). The QCRB is the ultimate estimation uncertainty bound with which a parameter ($\Delta\phi_S$) can be measured if all technical noise sources are eliminated, the measurements are completely ideal, and the states are perfectly prepared each time~\cite{Polino2020}. 

For a single-mode coherent state, the quantum Fisher information $\mathcal{I}_Q$ for phase estimation is calculated as~\cite{Xiao2019},
\begin{equation}\label{eq:qfish}
    \mathcal{I}_Q = \left(\frac{\partial d}{\partial \Delta\phi'}\right)^T \Sigma^{-1} \frac{\partial d}{\partial \Delta\phi'} + \Tr \left[ \left( \frac{\partial \Sigma}{\partial \Delta\phi'} \Sigma^{-1} \right)^2 \right],
\end{equation}

\noindent where ${\Delta\phi' =  \Delta\phi_R -  \Delta\phi_S}$, $d$ represents the displacement vector, ${d = \frac{1}{2}|\alpha_S|\{\cos \Delta\phi', -\sin \Delta\phi', \sin \Delta\phi', \cos \Delta\phi'\}^T}$,  and $\Sigma$ is the covariance matrix,
\begin{equation}\label{eq:cov}
    \Sigma_{ij} = \frac{1}{2} \langle \mathcal{X}_i \mathcal{X}_j + \mathcal{X}_j \mathcal{X}_i \rangle - \langle \mathcal{X}_i \rangle \langle \mathcal{X}_j \rangle,
\end{equation}

\noindent for the quadratures vector $\mathcal{X} = [X_S, P_S]$. The QCRB is then calculated as ${\mathrm{QCRB} \geq 1/(N \mathcal{I}_Q)}$. For a single-mode coherent state used for phase estimation, the QCRB becomes~\cite{Xiao2019},
\begin{equation}\label{eq:qcrb}
    \mathrm{QCRB} \geq \frac{1}{2N |\alpha_S|^2},
\end{equation}

\noindent where $|\alpha_S|$ is the signal amplitude, ${|\alpha_S| = \sqrt{X_{S}^2 + P_{S}^2}}$. 

\section{Neural Network Analysis}  \label{sec:nn_anal}
\subsection{Case 1}\label{sec:case_1}
Case~1 focuses on the Gaussian phase noise (described in Section~\ref{sec:rpe}), where the NN essentially learns to estimate $\Delta\phi_S$ by taking the mean of the previous $N$ measurements of $\Delta\phi_R$. 

The performance of the NN-based $\Delta\phi_S$ estimation is quantified by the estimation error variance between the estimate $\Delta\Tilde{\phi}_S$ and true $\Delta\phi_S$, ${\mathrm{Var}(\Delta\Tilde{\phi}_S - \Delta\phi_S)}$, where ${\mathrm{Var}(\Delta\Tilde{\phi}_S - \Delta\phi_S) = 0}$ would represent perfect $\Delta\phi_S$ estimation. When the approximation $\Delta\phi_R \approx \Delta\phi_S$ is applied, then the estimation error variance is ${\mathrm{Var}(\Delta\phi_R - \Delta\phi_S)}$, while for the expectation estimator, the estimation error variance is ${\mathrm{Var}(\langle \Delta\phi_R \rangle_N - \Delta\phi_S)}$.

Table~\ref{tab:nn_comp_c1} outlines the different NN architectures investigated for Case~1. $\#$ represents the assigned NN model numbers, with Model~0 representing the approximation $\Delta\phi_R \approx \Delta\phi_S$ (included for comparison). Taking the results of Model~3 as an example, ${\mathrm{Var}(\Delta\phi_R - \Delta\phi_S) = 0.1123}$ shot noise units (SNU) is reduced to ${\mathrm{Var}(\Delta\Tilde{\phi}_S - \Delta\phi_S) = 0.0033 \ \mathrm{SNU}}$ after the NN output $\Delta\Tilde{\phi}_S$ is applied to the RLO$_2$. The NN estimation performance matches the expectation estimator at ${\mathrm{Var}(\langle \Delta\phi_R \rangle_N - \Delta\phi_S) = 0.0033 \ \mathrm{SNU}}$, indicating that the NN is able to learn to take the mean of the previous $N$ measurements of $\Delta\phi_R$ for the Gaussian phase noise.

\begin{table}[htbp]
    \begin{center}
        \caption{Case~1 neural network architecture performance.}
        \label{tab:nn_comp_c1}
        \begin{tabularx}{\linewidth}{ 
            >{\raggedright\arraybackslash}s
            | >{\centering\arraybackslash}m
            | >{\centering\arraybackslash}m
            | >{\centering\arraybackslash}l
            | >{\centering\arraybackslash}e}
            \hline
            $\#$ & $N$ & $z_{dim}$ & $S$ & {Est. Error Variance} \\ \hline \hline
            0 & - & - & - & 0.1123 SNU \\ \hline 
            1 & 20 & 4 & 2,020 & 0.0055 SNU \\ \hline 
            2 & 20 & 32 & 87,700 & 0.0052 SNU \\ \hline 
            3 & 40 & 4 & 4,040 & 0.0033 SNU \\ \hline 
            4 & 60 & 4 & 6,060 & 0.0022 SNU \\ \hline 
            5 & 80 & 4 & 8,080 & 0.0016 SNU \\ \hline 
            6 & 100 & 4 & 10,100 & 0.0015 SNU \\ \hline 
            7 & 100 & 32 & 438,500 & 0.0014 SNU \\ 
        \end{tabularx}
    \end{center}
\end{table}

Fig.~\ref{fig:crb} shows the number of $\Delta\phi_R$ measurements $N$ versus ${\mathrm{Var}(\Delta\Tilde{\phi}_S - \Delta\phi_S)}$ for each of the Case~1 NN models, as well as the QCRB. The models are defined by their scale, with Case~1 results shown with solid markers, while the NN model Case~2 results are shown with hollow markers, with the expectation estimator results for Case~2 plotted as hollow blue star markers (discussed in Section~\ref{sec:case_2}). It can be seen that none of the $\Delta\phi_S$ estimation models are able to reach the QCRB given the imperfections of the receiver, though the results for $N=20$ approach it. The estimation error variance of the Case~1 NN models appear to follow the same trend as the QCRB. 

First we fix the number of $\Delta\phi_R$ measurements at ${N=100}$, then vary the number of hidden units from ${z_{dim}=4}$ to ${z_{dim}=32}$. The increased estimation performance of Model~6, compared to Model~7, is negligible, as featured in the Fig.~\ref{fig:crb}~inset~(b), while the scale increases from ${S=10,100}$ to ${S=438,500}$. The same range of $z_{dim}$ is also tested for ${N=20}$, with a slightly greater difference in estimation performance, as featured in Fig.~\ref{fig:crb}~inset~(a). The number of $\Delta\phi_R$ measurements is also varied between ${N=20}$ and ${N=100}$, fixing ${z_{dim}=4}$. Overall, it is shown that increasing the number of $\Delta\phi_R$ measurements has a greater increase in $\Delta\phi_S$ estimation performance than increasing $z_{dim}$ within each cell. 

The increase in $\Delta\phi_S$ estimation performance for Model~3 (outlined with the green oval in Fig.~\ref{fig:crb}), compared to Models~1-2, while only having a scale of ${S=4,040}$, is more significant than the increase in estimation performance for Models~4-7. As such, Model~3 presents a low-complexity $\Delta\phi_S$~estimation algorithm, which satisfies the performance requirements of a practical CV-QKD system. 

\subsection{Case 2}\label{sec:case_2}
In Case~2 we introduce an additional phase error component between $\Delta\phi_R$ and $\Delta\phi_S$, modeling a relationship between $\Delta\phi_R$ and $\Delta\phi_S$ that remains constant during each coherence time. It is assumed that the relationship between $\Delta\phi_R$ and $\Delta\phi_S$ could be represented by any function, where the LSTM should be able to map $\Delta\phi_R$ to $\Delta\phi_S$ for varying relationships. For demonstrative purposes, we introduce a Gaussian mixture model for the $\Delta\phi_R$ measurements using the additional PDF $\mathcal{N}(0.0000$, $0.0524$). The Gaussian phase noise from Case~1 still applies to every $\Delta\phi_R$ measurement. The same NN architectures as in Case~1 are tested for Case~2, outlined in Table~\ref{tab:nn_comp_c2}. Model~8 represents the approximation $\Delta\phi_R \approx \Delta\phi_S$.

\begin{table}[htbp]
    \begin{center}
        \caption{Case~2 neural network architecture performance.}
        \label{tab:nn_comp_c2}
        \begin{tabularx}{\linewidth}{ 
            >{\raggedright\arraybackslash}s
            | >{\centering\arraybackslash}m
            | >{\centering\arraybackslash}m
            | >{\centering\arraybackslash}l
            | >{\centering\arraybackslash}e}
            \hline
            $\#$ & $N$ & $z_{dim}$ & $S$ & {Est. Error Variance} \\ \hline \hline 
            8 & - & - & - & 0.1294 SNU \\ \hline 
            9 & 20 & 4 & 2,020 & 0.0065 SNU \\ \hline 
            10 & 20 & 32 & 87,700 & 0.0049 SNU \\ \hline 
            11 & 40 & 4 & 4,040 & 0.0050 SNU \\ \hline 
            12 & 60 & 4 & 6,060 & 0.0070 SNU \\ \hline 
            13 & 80 & 4 & 8,080 & 0.0065 SNU \\ \hline 
            14 & 100 & 4 & 10,100 & 0.0053 SNU \\ \hline 
            15 & 100 & 32 & 438,500 & 0.0062 SNU \\ 
        \end{tabularx}
    \end{center}
\end{table}

\begin{figure}
    \includegraphics[scale=0.87]{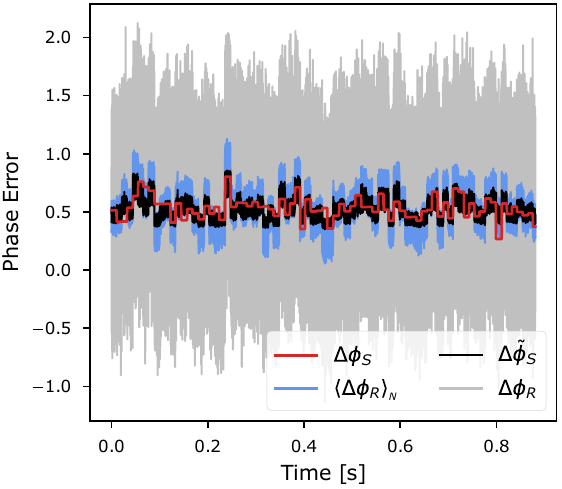}
    \caption{Signal phase error estimation results for Case~2 using a Model~11 example. The $\Delta\phi_R$ measurements are plotted as the grey line, the true $\Delta\phi_S$ values as a red line, the NN estimate $\Delta\Tilde{\phi}_S$ as a black line, and the expectation estimator output as the blue line.} \label{fig:lstm_est}
\end{figure}

Fig.~\ref{fig:lstm_est} gives an example of the NN output $\Delta\Tilde{\phi}_S$ for Model~11, with ${N=40}$ and ${z_{dim}=4}$ resulting in a scale of ${S=4,040}$. The $\Delta\phi_R$ measurements are shown as the grey line, the $\Delta\phi_S$ values as the red line, the $\Delta\Tilde{\phi}_S$ estimations by the black line, and the expectation estimator results by the blue line. The estimation performance of Model~11 reduces ${\mathrm{Var}(\Delta\phi_R - \Delta\phi_S) = 0.1294 \ \mathrm{SNU}}$ to ${\mathrm{Var}(\Delta\Tilde{\phi}_S - \Delta\phi_S) = 0.0050 \ \mathrm{SNU}}$. The expectation estimator results in ${\mathrm{Var}(\langle \Delta\phi_R \rangle_N - \Delta\phi_S) = 0.0145 \ \mathrm{SNU}}$, highlighting the advantage of our NN-based $\Delta\phi_S$~estimation algorithm over standard approaches when the relationship between $\Delta\phi_R$ and $\Delta\phi_S$ becomes more complex, as can be seen in Fig.~\ref{fig:lstm_est} by comparing the black line ($\Delta\Tilde{\phi}_S$) with the blue line ($\langle \Delta\phi_R \rangle_N$).

\begin{figure}
    \begin{centering}
    \includegraphics[scale=0.82]{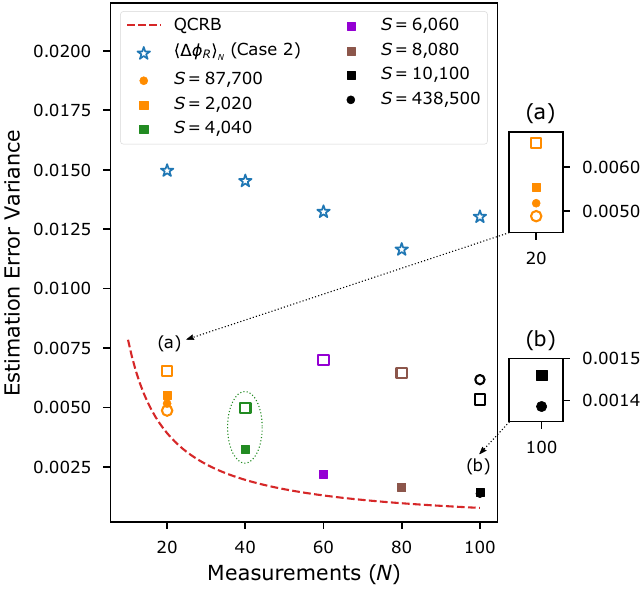}
    \caption{$N$ measurements of $\Delta\phi_R$ versus estimation error variance for NN architectures tested, defined by their scale $S$, as well as the QCRB. Case~1 NN results are shown with solid markers, Case~2 NN results are shown with hollow markers, and Case~2 expectation estimator $\langle \Delta\phi_R \rangle_N$ results are shown with hollow blue star markers. Inset~(a) highlights results for Models~1-2 and~9-10, and inset~(b) highlights the results for Models~6-7.}\label{fig:crb}
    \end{centering}
\end{figure}

Returning to Fig.~\ref{fig:crb}, the $\Delta\phi_S$ estimation performance trends differ for the Case~2 NN models than for Case~1, shown as the hollow markers. With the $\Delta\phi_R$ measurements fixed at ${N=20}$, there is a larger increase in estimation performance when hidden units are increased from ${z_{dim}=4}$ to ${z_{dim}=32}$, as shown in Fig.~\ref{fig:crb}~inset~(a). On the contrary, when the $\Delta\phi_R$ measurements are fixed at ${N=100}$, Model~14 ($z_{dim}=4$) actually performs better than Model~15 ($z_{dim}=32$). When the $\Delta\phi_R$ measurements are varied from ${N=20}$ to ${N=100}$, it can be seen that the estimation performance initially increases with $N$, before reducing above ${N=40}$. Thus, the NN is better able to learn the relationship between $\Delta\phi_R$ and $\Delta\phi_S$ for lower $N$. It is shown that the NN-based signal error estimation algorithm outperforms the expectation estimator (plotted as blue stars in Fig.~\ref{fig:crb}) across all values of $N$. 

After comparing different architectures for the NN-based $\Delta\phi_S$~estimation algorithms, we show that an efficient low-complexity model can be obtained. \textit{This is the main result of this work} - the conclusion that NNs an order of magnitude reduced in complexity can provide for suitable performance levels. For context, the previous work~\cite{Zhang2023} utilizes an LSTM for phase estimation, with an architecture of ${N=3}$ and ${z_{dim}=100}$, resulting in a scale of ${S=122,400}$. While it is assumed that each CV-QKD setup will have different phase error relationships, similar estimation performance could potentially be achieved with a lower complexity model. For example, our work shows a limited increase in $\Delta\phi_S$ estimation performance for the higher complexity models with the same value of $N$ (see Fig.~\ref{fig:crb} insets). For Case 2, the $\Delta\phi_S$ estimation performance of Model~11 outperforms the higher complexity Models~12-15, while achieving similar performance to Model~10, which has an order of magnitude greater scale. As such, our low-complexity NN-based $\Delta\phi_S$~estimation algorithm is capable of satisfying the requirements for a practical real-time CV-QKD system, improving the computational speed of the algorithm by reducing the NN scale, while achieving sufficient $\Delta\phi_S$ estimation. Our work reduces the computational requirements for phase estimation, which is essential for real-time CV-QKD.

\section{Conclusion} \label{sec:conc}

Our work analyzes LSTM-based neural network architectures for signal phase error estimation, designed for a satellite-to-Earth CV-QKD system. The results show that adequate signal phase error estimation can be achieved using a low-complexity LSTM NN architecture, with the NN-based approach outperforming standard models when the relationship between the reference pulse phase error and signal phase error became more complex. Decreasing the complexity of $\Delta\phi_S$~estimation algorithms remains vital for achieving real-time atmospheric CV-QKD. 

\printbibliography

\end{document}